# Studies on Magnetic Excitation Spectra of High-$T_c$ Superconductors

Masafumi Ito, Yukio Yasui, Satoshi Iikubo, Minoru Soda,
Akito Kobayashi and Masatoshi Sato

*Department of Physics, Division of Material Science, Nagoya University,
Furo-cho, Chikusa-ku, Nagoya 464-8602*

Kazuhisa Kakurai

*Advanced Science Research Center, JAERI, Tokai, Ibaraki 319-1195*

Chul-Ho Lee

*National Institute of Advanced Industrial Science and Technology, Tsukuba, Ibaraki
305-8568, Japan*

And

Kazuyoshi Yamada

*Institute for Materials Research, Tohoku University, Sendai, 980-8577, Japan*

**Abstract**

Magnetic excitation spectra $\chi''(q,\omega)$ of $YBa_2Cu_3O_y$, $La_{2-x}Sr_xCuO_4$, $La_{2-y-x}Nd_ySr_xCuO_4$ and $Nd_{2-x}Ce_xCuO_4$ are compared with those of model calculations, where $q$ and $\omega$ are the wave vector and the energy of the excitations. In the model, the dynamical spin susceptibility is calculated by the expression $\chi(q,\omega) = \chi_0(q,\omega)/\{1+J(q)\chi_0(q,\omega)\}$, where $\chi_0(q,\omega)$ is obtained by adopting the effective energy dispersion of the quasi particles which can reproduce the shapes of the Fermi surface, and $J(q)$ is the exchange coupling between the Cu-Cu electrons. It is important to consider the experimentally observed quasi particle broadening ($\Gamma$) to reproduce the observed spectra of $YBa_2Cu_3O_y$, while to reproduce the data of $La_{2-x}Sr_xCuO_4$ and $La_{2-y-x}Nd_ySr_xCuO_4$, very small $\Gamma$ values have to be used, even in the phases without the static "stripe" order. This may be caused by the existence of the "stripe" correlations or "stripe" fluctuations.

corresponding author: M. Sato (e-mail: msato@b-lab.phys.nagoya-u.ac.jp)

# 1. Introduction

Since the discovery of high-$T_c$ superconductors, the anomalous metallic phase of Cu oxides has attracted much attention to clarify the mechanism of high-$T_c$ superconductivity, and importance of the magnetic excitations is recognized for the realization of unusual behaviors observed in the phase.[1-4] One of effective ways to obtain information on magnetic characteristics of high-$T_c$ superconductors is to measure their magnetic excitation spectra $\chi''(q,\omega)$ by means of neutron inelastic scattering, where $q$ and $\omega$ are the wave vector and the energy of excitation, respectively.

In the first stage of the studies on $\chi''(q,\omega)$, attention has been focused on the pseudo gap behavior in the normal state of the underdoped $YBa_2Cu_3O_y$(YBCO or YBCO$_y$).[5,6] Such the pseudo gap structure can also be found in the electronic tunneling density of states even far above $T_c$.[7-9] It shows how the singlet correlation grows with decreasing temperature $T$. The existence of the resonance peak is also attractive phenomenon,[10,11] which shows detailed electronic structure of high-$T_c$ oxides. Because detailed structure of $\chi''(q,\omega)$ and its $T$ dependence is rather characteristic(The incommensurate (IC)-commensurate-IC variation with increasing $\omega$ observed below $T_c$ for YBCO[12] is one of these examples), it is possible to compare characteristic of model calculations with the observed ones.

The existence of sharp IC peaks of $\chi''(q,\omega)$ observed for La214 systems such as $La_{2-y-x}Nd_ySr_xCuO_4$ and $La_{2-x}Ba_xCuO_4$ even above $T_c$[13,14] are also important characteristic of the electron systems. It seems to be related with the formation of "stripes", which consist of one-dimensional chains of holes and antiferromagnetically ordered spins between the chains. The $T_c$ suppression found in La214 systems at around $x\sim 1/8$ (1/8 anomaly), is considered to be caused by the "stripe" ordering.[15,16] The "stripe" model has been proposed to explain the superlattice reflections at the IC points in the reciprocal space of $La_{2-y-x}Nd_ySr_xCuO_4$ with $x \sim 0.125$.[17-19] For this sample, superlattice peaks originated from the magnetic ordering appear at $q=(0.5\pm\delta,0.5)$ and $(0.5,0.5\pm\delta)$ (in tetragonal notation). If the slowly fluctuating (or dynamical) "stripes" exists, similar characteristics seen in $\chi''(q,\omega=0)$ of $La_{2-y-x}Nd_ySr_xCuO_4$ are also expected in the spectral function $\chi''(q,\omega)$ of other systems, and since similar IC peaks were observed in superconducting La214 system or YBCO,[20,21] dynamical "stripes" or the fluctuating "stripes" were suggested to play an important role in the occurrence of the high-$T_c$ superconductivity.[22]

Studies of electron-doped superconductors have also been carried out, in particular, to collect information of the electron-hole symmetry of high $T_c$ cuprates. The $\chi''(q,\omega)$ of $Nd_{2-x}Ce_xCuO_4$(NCCO) has been measured by Yamada *et al.* and commensurate peaks have been observed.[23] We expect that comparison of $\chi''(q,\omega)$ of several high-$T_c$ systems give further information on electronic nature of these systems.

Kao *et al.*[24] have first suggested that an expression

$$\chi(q,\omega)= \chi_0(q,\omega)/\{1+J(q)\chi_0(q,\omega)\} \tag{1}$$

can reproduce the IC-commensurate-IC variation of $\chi''(q,\omega)$ observed in YBCO with increasing $\omega$ below $T_c$, where $J(q)$ (=$J_0$(cos $q_x a$+cos $q_y a$), with $a$ being the lattice constant), is the exchange coupling between the Cu-Cu electrons. The wave vector $q$ is described by $q$=($q_x$, $q_y$,$q_z$) or ($q_x$, $q_y$) in this paper. When we use the Lindhart form of $\chi_0(q,\omega)$ in eq. (1), we just need the energy level distribution of electrons and the model is appropriate to the electron system irrespective of whether it is Fermi liquid-like or not.

We calculated $\chi''(q,\omega)$ in the similar way by using the effective energy dispersions of the quasi particles, and reproduced the IC-commensurate-IC variation of $\chi''(q,\omega)$ with increasing $\omega$.[25,26] In the calculation of $\chi_0(q,\omega)$, the effective band parameters $t_0$, $t_1$, $t_2$ are required to reproduce Fermi surface shape of the quasi particles, where the $t_0$, $t_1$ and $t_2$ have the usual meanings. Effects of the quasi particle broadening ($\Gamma$) have to be taken into account, because the values of $\Gamma$ estimated from the optical and the transport studies are very large.[27-31] $\Gamma$ depends on temperature $T$ and the quasi particle energy $\varepsilon$ (see Fig. 1), and the consideration of $\Gamma$ has been found to be important to reproduce the electronic tunneling density of states.[25]

In this paper, the magnetic excitation spectra of high-$T_c$ cuprates are discussed. In the analyses, we have paid attention to the effects of the quasi particle broadening on $\chi''(q,\omega)$ and tried to clarify how the effects of dynamical or fluctuating "stripes" appear in $\chi''(q,\omega)$. In our fittings of the calculated spectra $\chi''(q,\omega)$ to the experimental data of YBCO, we have obtained reasonable fittings. (One might think that the anisotropy of the neutron inelastic scattering profiles of $\chi''(q,\omega)$ reported in ref. 32 for YBCO$_{6.5}$ clearly indicated one-dimensional nature of $\chi''(q,\omega)$, and therefore that the calculations by the above model based on the two-dimensional picture cannot be rationalized. However, it is easily found that the observed anisotropy can be fitted well by the two-dimensional model by using a quite reasonable choice of the background level.) The similar model can reproduce the nature of $\chi''(q,\omega)$ observed in NCCO, too. On the other hand, we have not succeeded in reproducing $\chi''(q,\omega)$ of La214 system, for which small $\Gamma$ values, much smaller than those reported in other experimental studies[27-31] should be used. It exhibits a clear contrast to the case for YBCO and NCCO. Based on these results, we discuss how the electron systems can be described, and argue if the slowly fluctuating "stripes" commonly exist or not in all high-$T_c$ cuprates studied here.

## 2. Experiments

In our experiments, large crystals are required to obtain reasonably strong scattering intensities. Details of the samples of YBa$_2$Cu$_3$O$_{6.5}$ ($T_c$~52 K) and YBa$_2$Cu$_3$O$_{6.7}$ ($T_c$~62 K) were described in ref. 33. The oxygen number $y$ is controlled by choosing proper annealing temperatures, and can be estimated from the $c$-axis lattice parameters. Crystals of La$_{1.48}$Nd$_{0.4}$Sr$_{0.12}$CuO$_4$ were prepared by the traveling solvent floating zone method.[34]

Neutron measurements on the single crystals of YBa$_2$Cu$_3$O$_y$ and La$_{1.48}$Nd$_{0.4}$Sr$_{0.12}$CuO$_4$ were carried out on a triple axis spectrometer ISSP-PONTA installed at JRR-3M of JAERI in Tokai. Scattered neutron energy ($E_f$) was selected using the 002 reflection of Pyrolytic graphite (PG) for both the monochromator and the analyzer. A PG filter was placed after the sample to eliminate higher order contamination from the scattered beams. The horizontal collimations were 40'-40'-80'-80'. The effective vertical collimations were 80'-240'-480'-650'. They were mounted in an Al can (with He exchange gas) attached to the cold finger of a Displex-type closed-cycle refrigerator.

For measurements of $\chi''(q,\omega)$ on YBCO$_y$, a crystal was oriented with the [110] and [001] axes in the scattering plane. Profiles are taken by scanning $h$ of $q=(h,h,l)$ with the transfer energies $E$ being fixed at several values. The scattered neutron energy ($E_f$) is fixed at 14.7 meV in the measurements of $E \leq 15$ meV. For $E \geq 15$ meV, $E_f$ was fixed at 30.5 meV. Due to the existence of the bilayer structure of this system, the intensities of $\chi''(q,\omega)$ are modulated along the $l$ direction.[35] Only for $E<24$ meV, we can choose the spectrometer setting to measure at $q$ with $|l|\sim 2$, where the $q$-resolution width $\Delta q_h$ along the [110] direction (~0.08 A$^{-1}$) is not too large to observe the IC peak of $\chi''(q,\omega)$. For $E>24$ meV, we have to choose the value of $|l|\sim 5$, at which the second peak of $\chi''(q,\omega)$ modulated along $|l|$ exists, to satisfy the condition of the energy and momentum conservation.

For the case of La$_{1.48}$Nd$_{0.4}$Sr$_{0.12}$CuO$_4$, the crystal was oriented with the [100] and [010] axes in the scattering plane. In the experiments, six aligned crystals with a typical size of about 7 mm$\phi \times$30 mm were used. Studies of the reflections corresponding to the 100/010 nuclear reflections which indicate the occurrence of the transition from the low temperature orthorhombic (LTO1) phase with the space group B*mab* to the low-temperature tetragonal (LTT) phase via an intervening orthorhombic (LTO2) phase with the space group P*ccn* show that the structural phase transition from the LTO1→LTO2→LTT start at temperature $T_s$ (~69K) with decreasing $T$.[34] The superlattice peaks at $Q$=(2+2$\delta$,0) and (0.5,0.5-$\delta$) which are considered to be induced by charge and magnetic orderings, respectively, appear at almost equal temperatures, which we call $T_e$. The value of $\delta$ has turned out to be $\delta(\omega=0)=0.119\pm 0.0005$ r.l.u. The details of the sample characterization are shown in ref. 34. Profiles of the scattering intensity $S(q,\omega)$ were taken by scanning $k$ of $q=(h,k,0)$ with various fixed transfer energies $E$ ($\leq 12$ meV). The scattered neutron energy ($E_f$) is fixed at 14.7 meV. The values of the resolution width $\Delta q_h$ along [010] direction are ~0.07, 0.06, 0.04 and 0.05 A$^{-1}$ at $E$=2.5, 5.0, 8.0 and 12.0, respectively.

## 3. Model of the calculations

We have calculated $\chi''(q,\omega)$ without considering the effects of "stripes". Detailed description of $\chi_0(q,\omega)$ can be found in ref. 36. Effective band parameter $t_0$, $t_1$ and $t_2$ are used. The ratios of these parameters are chosen to reproduce the shape of the Fermi surface. We set the ratios of the band parameters ($t_0$:$t_1$:$t_2$) to be 1:-1/6:1/5 for YBa$_2$Cu$_3$O$_y$ and Nd$_{2-x}$Ce$_x$CuO$_4$.[37,38] For La214 system, the ratios are set to be 1:-1/6:-1/15. Although $t_2/t_0$ is different from the usual value(=0), it does not affect the shape of the Fermi surface significantly. (The most sharp IC peaks are obtained by using the ratios 1:-1/6:-1/15.) The intra-atomic Coulomb interaction energy $U$ is set to be 0, and the strong correlation effects are taken into account by adopting a small effective value of $t_0$. The band width obtained by this $t_0$ is consistent with the value of the coherent in-gap band deduced by the $d$-$p$ model [36] or the holon band of the $t$-$J$ model.[37]

In our calculation, the broadening $\Gamma(\varepsilon)$ is simply assumed to be isotropic and a model which roughly reproduces characteristics of the experimentally observed $T$- and $\varepsilon$-dependence of $\Gamma$ is used. The model of $\Gamma(\varepsilon)$ is shown for YBCO$_{6.5}$ in Fig. 1 for example, where $\Gamma_0$ and $\Gamma_h$ are the $\Gamma$ values at the chemical potential $\mu$ and at $|\varepsilon|>2\Delta_0$, respectively. ($\Gamma(\varepsilon)=\Gamma_0+(\Gamma_h-\Gamma_0)$ F($\varepsilon$), where F($\varepsilon$) has a polynomial form up to the cubic order of $\varepsilon$. F($\varepsilon$) satisfies F(0)=0 and F(2$|\Delta_0|$)=1. ) The value of $\Gamma_h$ was assumed to be independent of $T$, while $\Gamma_0$ increases with increasing $T$ (see the inset of Fig. 1.).

The value of $\mu$ is determined to adjust the $q$-position or the incommensurability $\delta$ of the observed IC peak of $\chi"(q,\omega)$ at low temperatures and low energies. The $d$-wave form of the energy gap $\Delta_s(k)=(\Delta_0/2)(\cos k_x a - \cos k_y a)$ is also used. For underdoped YBCO systems, the amplitudes of the superconducting gap and pseudo gap above $T_c$ are taken to be equal and $T$-independent in the present $T$ region, by considering the pseudo gap survives up to very high temperatures, while in the overdoped region, the amplitude of the gap above $T_c$ is chosen to be zero. The values of $2|\Delta_0|$ is chosen to be similar to those observed in the electronic tunneling density of states. The exchange coupling $J_0$ was so chosen that the resonance peak energy observed in $\chi"(q,\omega)$ can be reasonably reproduced. (The resonance energy is defined as the energy at which the peak of $\chi"(q,\omega)$ is commensurate (at $q$=(1/2,1/2) $\equiv q_{AF}$) and its intensity is the largest as a function of $\omega$.

In the numerical calculations of $\chi_0(q,\omega)$ and $\chi(q,\omega)$, the two-dimensional $q_x$-$q_y$ plane of the reciprocal space ($|q_x|,|q_y| < \pi/a$. Hereafter, the indices of $x$ and $y$ correspond to the $a$-, $b$-axes of the sample.) is divided into 100×100 cells. For the energy integration of the Green function, the energy step of the integration is taken to be 0.5 meV, which is always smaller than the broadening $\Gamma(\varepsilon)$. The region of the energy integration is 1 eV, which is larger than the band width ($8t_0$). The $t_0$ values are chosen to be proportional to the carrier concentration ($p$).

In the following, we just show how the choices of the parameters stated above affect the calculated characteristics of $\chi"(q,\omega)$. Figure 2(a) shows the energy dependence of $\chi"(q_{AF},\omega)$ calculated at $T$=7 K for various values of $J_0$ and other fixed parameters $t_0$=−20, $2\Delta_0$=88, $\mu$=0, $\Gamma_0$=4 and $\Gamma_h$= 50 meV, which seem to be reasonable for YBa$_2$Cu$_3$O$_{6.5}$. With increasing $J_0$, the peak height increases and the resonance energy decreases. Figure 2(b) shows the resonance energy ($\omega_{res}$) obtained as a function of $J_0$ for the same values of the other fixed parameters as those of Fig. 2(a). The $\delta$ value increases with decreasing $\mu$. In Figs. 3(a) and 3(b), the calculated results of the energy dependence of $\chi_0'(q,\omega)$ at $T$=7 K are shown. Open and filled circles in Fig. 3(a) are obtained for $\Gamma_h$= 4 and 50 meV, respectively, where other parameters are $t_0$=–20, $2\Delta_0$~0, $\mu$=0 and $\Gamma_0$=$\Gamma_h$. Open and filled circles in Fig. 3(b) are obtained for $\Gamma_h$= 4 and 50 meV, respectively, where other parameters are $t_0$=–20, $2\Delta_0$~88, $\mu$=0 and $\Gamma_0$=4 meV. We can see that the suppression of $\chi_0'(q,\omega)$, which indicates the suppression of antiferromagnetic tendency, is caused by the effects of quasi particle broadening and/or the superconducting gap. We have chosen the physically reasonable values of these parameters and tried to reproduce the observed $\chi''(q,\omega)$.

## 4. Results and Discussion

In Fig. 4, magnetic excitation spectra $\chi''(q,\omega)$ calculated for the underdoped YBCO$_{6.5}$ at 7 K are shown along ($h$,1/2) in the 2D reciprocal space. In the calculation, we used following parameters: $t_0$=-20 meV, $2\Delta_0$=88 meV, $J_0$=58.5 meV and $\mu$=0 meV. The $J_0$ value used here is roughly consistent with the experimental value estimated by Johnston.[39] The broadening $\Gamma(\varepsilon)$ (full width at half maximum) shown in Fig. 1 is used in the calculation. The inset of Fig. 4 shows the $q$–position at which IC and commensurate peaks appear. One can see that the IC-commensurate-IC variation with increasing $\omega$ is reproduced even in the case of the large $\Gamma$. The commensurate peak (or the resonance peak) appears at the energies of ~28 meV, which is consistent with the experimental data.[40, 41]

The solid lines in Fig. 5(a) show the result of the simultaneous fitting of the curves obtained after convoluting the resolution function with the calculated profiles of $\chi''(q,\omega)$ shown in Fig. 4, to the experimental data at 7 K with the scale factor as a common fitting parameter. (Detailed descriptions of the calculations including the resolution convolution can be found in refs. 25 and 26.) Broken lines are the results for the smaller values of $\Gamma_h$ (=4 meV) obtained in the similar way to the case of $\Gamma_h$= 50 mev. It should be noted that, by using the $\Gamma_h$ value of 50 meV, which is, roughly speaking, as large as experimentally observed $\Gamma_h$,[27-31] we obtain reasonable fittings to the data taken by scanning the scattering vector $q$ along ($h$, $h$, –2). Figure 5(b) shows the observed $\chi''(q,\omega)$ at 147 K together with the results of fittings to the experimental data. The fittings have been carried out in the similar way to the case of $T$=7 K. The solid lines have been obtained by using the same values of $t_0$, $2\Delta_0$, $J_0$, $\mu$ and $\Gamma_h$ as in Fig. 4 except $\Gamma_0$ and $T$, and the broken lines have been obtained by using the same values of $t_0$, $2\Delta_0$, $J_0$ and $\mu$ as in Fig. 4 except $\Gamma_0$, $\Gamma_h$ and $T$. The scale factors for the solid and broken lines are equal to the values used in the calculations of the corresponding lines in Fig. 5(a), respectively. $\Gamma_0$=40 meV and $\Gamma_h$= 50 meV are used in the calculation of the solid lines in Fig. 5(b), and for the broken lines, $\Gamma_0$=$\Gamma_h$=4 meV. Only the large value of $\Gamma_h$ can successfully reproduce the experiment data at 147 K. We stress that the reasonable fittings have been obtained by considering the effects of quasi particle broadening shown in Fig. 1.

Figure. 6(a) shows the ω-dependence of the $q$-dependent part of χ"($q$,ω) at $q$=(1/2,1/2) in the reciprocal space observed at 7K and 147 K. The data observed by Bourges et al.[42] are also shown by the open circles. Calculated results at 7 K and 147 K are shown by the solid and broken lines, respectively. Figures 6(b) and 6(c) show the ω-dependence of the $q$-dependent part of χ"($q$,ω) at (1/2,1/2) in the reciprocal space observed at 4.5 K and 95 K, respectively.[43] These data are fitted by using a common scale factors and following parameters. $t_0$=−25, $J_0$=56, μ=−5 and $\Gamma_h$=40 meV. At 4.5 K, $\Gamma_0$=4 meV and |$2\Delta_0$|=60 meV and at 95 K, $\Gamma_0$=40 meV and |$2\Delta_0$|=0. The fitted results are shown by the solid lines. Our model seems to reproduce the $q$–, ω–, $T$– and $y$– dependences of the magnetic excitation spectra of YBCO$_y$. In our calculation, quasi particle broadening has been found to be important to suppress the tendency of the antiferromagnetic ordering. The introduction of large Γ smears out the density of states around the Fermi surface, as is understood by the expression of the quasi particle propagator G~1/(ω-ε+iΓ), which causes the suppression of $\chi_0$($q_{AF}$,ω), $q_{AF}$ being the point (1/2,1/2) in the reciprocal space. (In the actual numerical calculations, the energy broadening Γ was assumed to be a function of ε.)

The results presented above indicate that the model can explain detailed structures of χ"($q$,ω) of YBCO system well, without considering the effects of dynamical "stripes", at least up to the energy of ~25 meV. Then, to clarify the effects of "stripes", we deliberately adopted La$_{1.48}$Nd$_{0.4}$Sr$_{0.12}$CuO$_4$, which has the static "stripe" ordering below $T_e$(~69 K). We have studied the detailed $T$ dependences of χ"($q$,ω) at 2.5 and 5.0 meV. The details for the measurements are shown in ref. 34.

As in the case of YBa$_2$Cu$_3$O$_y$, our calculations indicate the existence of the IC–commensurate–IC variation of the peak structure of χ"($q$,ω) in La214 systems, too, with increasing ω. In the actual experiments, however, it is not clear whether the commensurate (or resonance) peak exists in these systems, because the reliable data have not been taken up to the corresponding energy ω$_{res}$(≥20 meV). Figures 7(a) and 7(b) show the observed $S$($q$,ω)≡($n$+1)χ"($q$,ω) of La$_{1.48}$Nd$_{0.4}$Sr$_{0.12}$CuO$_4$ at ω=2.5 and 5 meV, respectively, together with the fitted curves, where $n$ is the Bose factor. The solid lines have been obtained by the fitting to the profiles with a common scale factor as a fitting parameter. Other quantities are $t_0$=−40 meV, $2\Delta_0$=0 meV, $J_0$=55 meV and $\Gamma_h$=$\Gamma_0$=2 meV. (The calculated curves have wavy structures by the coarseness of the integration.) The profiles at $T$=63 K in Fig. 7(a) and at $T$=66 K in Fig. 7(b), where the static "stripes" are almost condensed, cannot be reproduced even by using a quite small value of the broadening $\Gamma_h$ (=$\Gamma_0$). The peak widths of the calculated profiles below $T_e$ are usually much broader than those observed by the experiment, because the static order of the moments is not expected in the calculation. Even above $T_e$, the $q$ widths of the calculated profiles at 111 K (Fig. 7(a)) and at 122 K (Fig. 7(b)) are broader than those of the observed ones (again, even for the very small value of $\Gamma_h$). The broken lines in these figures, show the results of the fittings by using a large value of $\Gamma_h$(=50 meV). For this value of $\Gamma_h$, another common scale factor is used. The calculated profile widths are quite large, and even IC peaks are not reproduced. Thus, we have found that as long as we use the model of $\Gamma(\varepsilon)$ shown in Fig. 1, $\Gamma_h$ has to be much smaller than the values expected from other experimental data[30,31] reported for several high-$T_c$ oxides, to obtain the reasonable fittings.

In Fig. 8, the energy dependence of $\chi''(q,\omega)$ observed at $T$=83 K is shown together with the calculated results. The fitting parameters are equal to those of the solid lines of Figs. 7(a) and 7(b) except the value of $T$. (Note that again the very small values of $\Gamma_0=\Gamma_h$=2 meV are used.) The calculated results can roughly reproduce the nature of the energy dependence of $\chi''(q,\omega)$.

In Fig. 9, the temperature dependence of the profile widths (full width at half maximum) of $\chi''(q,\omega)$, $\Delta q$ of $La_{1.48}Nd_{0.4}Sr_{0.12}CuO_4$ is shown by filled and open circles for $\omega$=2.5 and 5.0 meV, respectively. They are estimated by fitting a double Gaussian line to the observed data (resolution corrected). The solid lines are guides for the eyes. The values of $\Delta q$ exhibit a significant decrease with decreasing $T$ through $T_s$ and becomes very small at low temperatures.

Now, we have shown that the data of $\chi''(q,\omega)$ taken for $La_{1.48}Nd_{0.4}Sr_{0.12}CuO_4$ cannot be reproduced by the calculations which do not consider possible effects of "stripes". It is in clear contrast to the result for the YBCO system. In particular, the rather small widths of the IC peaks in the high temperature region can hardly be reproduced. Is it due to certain effects of the slowly fluctuating "stripes" in $La_{1.48}Nd_{0.4}Sr_{0.12}CuO_4$? To answer the question, we consider what effects fluctuating "stripes" bring about on the behavior of $\chi''(q,\omega)$: If the slowly fluctuating "stripes" grow, the $CuO_2$ plane may be divided into two regions, hole-poor and hole-rich regions. The hole-poor region in the "stripes" mainly contributes to magnetic scattering. Due to the enhanced antiferromagnetic correlation in the hole-poor region, the width of the IC peak seems to become sharper than that expected from the calculations of $\chi''(q,\omega)$ described above. On the other hand, the broadening observed by other kinds of measurements is the one determined mainly from the hole-rich part. It seems to be able to explain why the broadening $\Gamma$ of only the spin part is so small. Then, the decrease in $\Delta q$ in a rather wide $T$ region as $T$ approaches $T_s$ from high $T$ side can be considered to be due to the growth of "stripe" fluctuations. The behavior of $\Delta q$ above $T_s$ (or $T_e$) seems to be proportional to $(T-T_e)^{1/2}$. These results directly indicate that the slowly fluctuating stripes exist even above $T_e$ (or $T_s$). However, it is not easy to distinguish if slowly fluctuating "stripes" exist in the ideal LTO1 phase or their existence is just due to the existence of the local LTT or LTO2 phase.

Next, we discuss the spectral function $\chi''(q,\omega)$ of $La_{2-x}Sr_xCuO_4$ system[14,44-46], in particular, to clarify if effects of dynamical "stripes" are seen or not in this system, which exhibits the superconducting transition and does not have, at least, the static order of the "stripes". In the analyses of data of $\chi''(q,\omega)$ reported by Lee et al.[44,45] (for $x$=0.10, 0.15 and 0.18), we apply a model which is similar to the one used for YBCO$_y$. For the present system, the effective band parameters $t_0$ of $x$= 0.10, 0.15 and 0.18 are chosen to be –30, –40 and –50 meV, respectively. The parameter $J_0$ is chosen to be 55 meV and the chemical potential $\mu$ is chosen to reproduce the $\delta$ values observed at low temperatures. We set the superconducting gap parameter $2\Delta_0$ to be 10~16 meV, considering the values observed in the STS measurements.[47] $\Gamma_0$ is always fixed at 4 meV. The broadening $\Gamma_h$ above $2\Delta_0$ is chosen to obtain reasonable fittings.

Figures 10(a)–10(c) show the results of the fittings to the profiles $\chi''(q,\omega)$ observed at several energies at low temperatures ($T$<10 K) for $x$=0.10, 0.15 and 0.18, respectively. For each $x$, a common scale factor is used to the set of the profiles. For $x$=0.10, we have used the following values of $t_0$=-30 meV, $2\Delta_0$=16 meV and $J_0$=55 meV. To obtain the reasonable fittings shown in Fig. 10(a), we have to use $\Gamma_h$=6 meV, which is much smaller than that used for YBCO$_y$ ($\Gamma_0$ is always fixed at 4 meV.). If we adopt $\Gamma_h$>6 meV, the calculated profiles become broader than the observed ones and the intensities also become too large at $q_{AF}$ to explain the observations. For $x$=0.15, $t_0$=-40 meV, $2\Delta_0$=15 meV, $J_0$=55 meV are used and $\Gamma_h$=16 meV is obtained. For $x$=0.18, $t_0$=-50 meV, $2\Delta_0$=10 meV, $J_0$=55 meV are used and $\Gamma_h$=20 meV is obtained. By using this parameter set for $x$=0.18, we have also calculated $\chi''(q,\omega)$ at several temperatures up to 150 K ($\omega$=5 meV) and found that the results can be fitted to the observed $T$ dependence of $\chi''(q,\omega)$ rather well by using a common scale factor and just by changing $\Gamma_0$ (6, 8 and 18 meV at 45, 80 and 150 K, respectively). In Fig. 9, we just show, as a function of $T$, the widths $\Delta q$ of the profiles of the above calculations and the observed ones for $x$=0.18 by the broken line and diamonds, respectively. These widths have been estimated by fitting the double Gaussian line to the observed and calculated profiles of La$_{1.82}$Sr$_{0.18}$CuO$_4$. One can know that the comparison between these two kinds of $\Delta q$ indicates that the $T$ dependent profiles $\chi''(q,\omega)$ can be explained by the above parameters well. The peak widths estimated for La$_{1.48}$Nd$_{0.4}$Sr$_{0.12}$CuO$_4$ are sharper than those of La$_{1.82}$Sr$_{0.18}$CuO$_4$.

Calculations of $\chi''(q,\omega)$ have also been carried out for $x$=0.06, 0.07, 0.08, 0.12 and 0.14 at $\omega$=2~3.5 meV to compare the results with the existing profiles to obtain information of $\Gamma_h$ in a wide region of $x$. They were taken at $T$=$T_c$ (12~32 K)[46] for $x$=0.06, 0.07, 0.08 and 0.12 and at $T$=20 K for $x$=0.14.[14] No well defined peaks of the magnetic reflection which indicate the existence of the static order at the incommensurate points at $q$=(0.5±$\delta$,0.5) and (0.5,0.5±$\delta$) were found at these temperatures. The effective band parameters $t_0$ of $x$=0.06, 0.07, 0.08, 0.12 and 0.14 are chosen to be –24, –25.5, –27, –35 and –38 meV, respectively. We set $2\Delta_0$ to be 16 and 15 meV for $x$=0.12 and 0.14, respectively, as is observed in the STS measurements.[47] For $x$<0.10, calculations have been carried out with various values of the superconducting gap $2\Delta_0$ in the region of 0~18 meV, and found that the choice of the $2\Delta_0$ value does not have significant effects to the results. $\Gamma_0$ is always fixed at 4 meV except $x$ =

0.12. (We have to use $\Gamma_0 = \Gamma_h = 3$ meV to reproduce sharp peaks observed for $x = 0.12$.) Then, the broadening $\Gamma_h$ above $2\Delta_0$ has been chosen to reproduce the observed profile.

The carrier concentration dependences of the quasi particle broadening ($\Gamma_h$) for YBCO$_y$, La$_{2-x}$Sr$_x$CuO$_4$ and La$_{2-y-x}$Nd$_y$Sr$_x$CuO$_4$ are shown in Fig. 11. We find that $\Gamma_h$ of La214 are much smaller than those of YBCO system. The difference also exists in the $p$ dependence of $\Gamma_h$ between these two systems: $\Gamma_h$ of YBCO decreases with increasing $p$ (or $y$), which is consistent with the results of the optical measurements.[31] Similar results are also reported in Bi2212 system.[31,48] In contrast, the suppression of $\Gamma_h$ in La$_{2-x}$Sr$_x$CuO$_4$ is found to be the most significant in the vicinity of $x \sim 1/8$. It may be understood by considering that the suppression of $\Gamma_h$ is driven by the slowly fluctuating "stripes".

We have also calculated $\chi''(q,\omega)$ of Nd$_{1.85}$Ce$_{0.15}$CuO$_4$, to see if our model is appropriate for the electron doped high-$T_c$ cuprates. In Fig. 12, the results of the calculation are shown, where we used $t_0(=-40$ meV), $t_1=-t_0/6$, $t_2=t_0/5$, $2\Delta_0=10$ meV, $J_0=55$ meV, $\Gamma_0=4$ meV and $\Gamma_h=35$ meV. We have chosen the value of the chemical potential $\mu=25$ meV, so that the electron number becomes 1.15. For these parameters, we can roughly reproduce the commensurate peak observed experimentally (See the inset figure of Fig. 12.)[23]

We have shown that our model, which uses the effective band parameters, exchange coupling between the Cu electrons and quasi particle broadening can well reproduce the experimentally observed characteristics of $\chi''(q,\omega)$ of YBCO$_y$ and NCCO systems. In particular, for YBCO system, we can reproduce the detailed $q$-, $\omega$-, $T$- and $y$- dependence of $\chi''(q,\omega)$, including the behavior of resonance peak. (Although $\Gamma_h$ is not strictly equal to the values estimated by other experimental studies, it may be due to the experimental uncertainties caused by various kinds of reasons. For example, the broadening is often estimated by the surface sensitive experiments, in which the value tends to be overestimated. It may also be due to incompleteness of the present model calculation.) By the present analyses, we can obtain new information on the electronic behaviors: To reproduce the observed characteristics (in particular, the profile width) of $\chi''(q,\omega)$ found for La214 system, we have to use small values of $\Gamma_h$ as compared to those observed by other methods, which seems to be understood by introducing the effects of the slowly fluctuating "stripes". Moreover, we have found that the suppression of $\Gamma_h$ is strong in La$_{1.48}$Nd$_{0.4}$Sr$_{0.12}$CuO$_4$ and in La$_{2-x}$Sr$_x$CuO$_4$ in the vicinity of $x \sim 1/8$. Even in the samples of La$_{2-x}$Sr$_x$CuO$_4$ for $x=0.15$ and 0.18, the broadening is significantly smaller than that of YBCO with the corresponding $p$ value. It indicates that the slowly fluctuating "stripes" survive up to such the large carrier concentrations in La214 system, and suggests that the effects of slowly fluctuating "stripes" exist only in the La214 system. We suppose that the "stripes" are connected with the structural characteristic of La214 system.

In the present study on La$_{1.48}$Nd$_{0.4}$Sr$_{0.12}$CuO$_4$, the effects of "stripes" are observed both in LTT and LTO1 phases. However, as was reported by Han *et al.*, the local structure of the LTO1 phase is not the ideal LTO1 structure but is similar to that of the LTO2 or LTT one even at room temperature,[49] and we cannot exclude a possibility that the persistence of the local structural characteristics of the low-temperature phase[49–51] may induce the short-range "stripe" correlation even in the temperature range of LTO1 phase.

Finally, we make a brief comment on the one-dimensional nature of $\chi''(q,\omega)$ reported by Stock *et al.*[32] we have tried to fit the results of our calculated profiles after the resolution convolution to their experimental data and obtained rather satisfactory fit just by using a reasonable choice of the background level. The anisotropy of the experimental data can be understood by considering the resolution effect. It indicates that the authors cannot extract the conclusion that $\chi''(q,\omega)$ has the one dimensional nature. We also note that by observing anomalous behavior of high energy phonon ($\omega \sim 60$ meV) in YBCO, the significance of the "stripes" is often argued.[52]. Our experiments on $\chi''(q,\omega)$ is just up to the energy <30 meV and therefore we cannot discuss if effects of "stripes" exist or not in such the high energy (or very short time scale) region. However, it should be noted that there exist very puzzling results reported by Pintschovius *et al.*[52] on YBCO that anomalous behavior observed on the high energy phonon indicates that the "stripes" should be along the *a*-axis even if it can be explained as their effects.

## 4. Summary

Magnetic excitation spectra $\chi''(q,\omega)$ of $YBa_2Cu_3O_y$, $La_{2-x}Sr_xCuO_4$, $La_{2-y-x}Nd_ySr_xCuO_4$ and $Nd_{2-x}Ce_xCuO_4$ are compared with those of model calculations. It is important to consider the experimentally observed quasi particle broadening ($\Gamma$) to reproduce the observed spectra of $YBa_2Cu_3O_y$, while to reproduce the data of $La_{2-x}Sr_xCuO_4$ and $La_{2-y-x}Nd_ySr_xCuO_4$, very small $\Gamma$ values, much smaller than those reported in other experimental studies have to be used, even in the phases without the static "stripe" order. The discussion on the relationship between the observed results and the "stripe" fluctuations has been presented.

## 5. Acknowledgements

The present work is supported by a grant-in-Aid for Scientific Research on Priority Areas from the Ministry of Education, Culture, Sports, Science and Technology of Japan. Two of the authors (M. I and S. I) are supported by a Research Fellowship of the Japan Society for the Promotion of Science for Young Scientists.

**Figure captions**

Fig. 1  The model function of $\Gamma(\varepsilon)$ used in the present calculations of $\chi''(q,\omega)$ of YBCO$_{6.5}$ are shown at several temperatures. $\Gamma_0$ and $\Gamma_h$ correspond to the $\Gamma$ values at $\varepsilon=0$ and $|\varepsilon|>2\Delta_0$, respectively. Inset shows the temperature dependence of $\Gamma_0$.

Fig. 2  (a) Energy dependence of $\chi''(q,\omega)$ at $q_{AF}=(1/2,1/2)$ calculated by using several $J_0$ values. (b) $J_0$ dependence of the $\omega_{res}$ is shown.

Fig. 3  (a) Energy dependence of $\chi_0'(q,\omega)$ calculated for YBCO$_{6.5}$ at 7 K for the effective band parameters $t_0(=-20$ meV$)$, $t_1=-t_0/6$, $t_2=t_0/5$, $U=0$ and $2\Delta_0\sim0$ meV. Open and filled circles represent the data obtained for the small quasi particle broadening ($\Gamma_0=\Gamma_h=4$ meV) and for the large quasi particle broadening ($\Gamma_0=50$ meV, $\Gamma_h=50$ meV). (b) Energy dependence of $\chi_0'(q,\omega)$ calculated for YBCO$_{6.5}$ for the effective band parameters $t_0(=-20$ meV$)$, $t_1=-t_0/6$, $t_2=t_0/5$ and $U=0$ and $2\Delta_0\sim88$ meV. Open and filled circles represent the data obtained for the small quasi particle broadening($\Gamma_0=\Gamma_h=4$ meV) and for the large quasi particle broadening ($\Gamma_0=4$ meV, $\Gamma_h=50$ meV).

Fig. 4  Magnetic excitation spectra $\chi''(q,\omega)$ calculated for YBCO$_{6.5}$ at 7 K by eq. (1) for the effective band parameters $t_0(=-20$ meV$)$, $t_1=-t_0/6$, $t_2=t_0/5$ and $U=0$, are shown along [10] direction in the 2-dimensional reciprocal space for several $\omega$ values. Inset figure shows the $q$ points, at which the incommensurate or commensurate peak appears.

Fig. 5  (a) Neutron Scattering intensities of the magnetic excitations taken at 7 K for YBCO$_{6.5}$ by scanning along $(h,h,\sim-2)$ at several fixed $\omega$ values (solid circles). Profiles obtained by convoluting the resolution functions with the calculated $\chi''(q,\omega)$ by using small $\Gamma(\Gamma_0=\Gamma_h=4$ meV$)$ and large $\Gamma(\Gamma_0=4$ meV, $\Gamma_h=50$ meV$)$ are also shown by the broken and solid lines. The zeros of the vertical axis are shifted upwards by 100, 100 counts/14400 kmon. for $\omega=15.0$ and 24.0 meV, respectively. (b) Results at 147 K are shown similarly to the case of Fig. 5(a). Profiles obtained by convoluting the resolution functions with the calculated $\chi''(q,\omega)$ by using small $\Gamma(\Gamma_0=\Gamma_h=4$ meV$)$ and large $\Gamma(\Gamma_0=40$ meV, $\Gamma_h=50$ meV$)$ are also shown by the broken and solid lines. The other parameters except $T$ and $\Gamma$ are chosen to be equal to those used in Fig. 5(a). The zeros of the vertical axis are shifted upwards by 200, 250 counts/14400 kmon. for $\omega=15.0$ and 24.0 meV, respectively.

Fig. 6  (a) Experimental and calculated peak intensities of the magnetic scattering at $q_{AF}=(1/2,1/2)$ in the two dimensional reciprocal space are shown at 7 K and 147 K, where the $q$ independent part of the intensities is removed. The data shown by the open circles are the values reported by Bourges *et al.* in ref. 42. (b),(c) experimental results of $\chi''(q,\omega)$ at $q_{AF}=(1/2,1/2)$ are shown together with the calculated ones, at 4.5 K and 95 K, respectively. Experimental data are from ref. 43.

Fig. 7   (a) Neutron Scattering intensities $S(q,\omega)$ of the magnetic excitations taken at $\omega=2.5$ meV for La$_{1.48}$Nd$_{0.4}$Sr$_{0.12}$CuO$_4$ by scanning along $(0.5,k,0)$ at several temperatures $T$ (solid circles). Profiles of the calculated $S(q,\omega)$ with $\Gamma(=\Gamma_0=\Gamma_h)=2$ and 50 meV are also shown by the solid and broken lines, respectively. The zeros of the vertical axis are shifted upwards by 150, 300, 450 counts/19500 kmon. for $T$=92.0, 75.0 and 63.0 K, respectively. (b) Neutron Scattering intensities $S(q,\omega)$ of the magnetic excitations taken at $\omega=5.0$ meV for La$_{1.48}$Nd$_{0.4}$Sr$_{0.12}$CuO$_4$ by scanning along $(0.5,k,0)$ at several temperatures $T$ (solid circles). Profiles of the calculated spectra $\chi"(q,\omega)$ with $\Gamma(=\Gamma_0=\Gamma_h)=2$ and 50 meV are also shown by the solid and broken lines, respectively. The zeros of the vertical axis are shifted upwards by 150, 220, 460 counts/19500 kmon. for $T$=92.0, 76.0 and 66.0 meV, respectively.

Fig. 8   Magnetic excitation spectra taken at $T$=83 K for La$_{1.48}$Nd$_{0.4}$Sr$_{0.12}$CuO$_4$ by scanning along $(0.5,k,0)$ at several energies (solid circles). The spectra $\chi"(q,\omega)$ calculated for the parameters $t_0=-40$ meV, $2\Delta_0\sim 0$ meV, $J_0=55$ meV, $\Gamma(=\Gamma_0=\Gamma_h)=2$ meV are also shown by the solid lines.

Fig. 9   Temperature dependences of the peak widths $\Delta q$ (FWHM) of the magnetic excitation spectra taken along the $k$ direction. Filled and open circles represent the data points obtained at $\omega=2.5$ and 5.0 meV on La$_{1.48}$Nd$_{0.4}$Sr$_{0.12}$CuO$_4$, respectively. The solid lines are guides for the eye. Filled diamonds represent the data points obtained at $\omega=5.0$ meV on La$_{1.82}$Sr$_{0.18}$CuO$_4$. The broken line shows the calculated results obtained by using $t_0=-50$ meV, $2\Delta_0=10$ meV, $J_0=55$ meV and $\Gamma_h=20$ meV.

Fig. 10   (a) Magnetic excitation spectra taken at $T$=9 K for La$_{1.90}$Sr$_{0.10}$CuO$_4$ by scanning along $(0.5,k,0)$ at several energies (solid circles). The spectra $\chi"(q,\omega)$ calculated for the parameters $t_0=-30$ meV, $2\Delta_0=16$ meV, $J_0=55$ meV, $\Gamma_0=4$ and $\Gamma_h=6$ meV are also shown by the solid lines. (b) Magnetic excitation spectra taken at $T$=8 K for La$_{1.85}$Sr$_{0.15}$CuO$_4$ by scanning along $(0.5,k,0)$ at several energies (solid circles). The spectra $\chi"(q,\omega)$ calculated for the parameters $t_0=-40$ meV, $2\Delta_0=15$ meV, $J_0=55$ meV, $\Gamma_0=4$ and $\Gamma_h=16$ meV are also shown by the solid lines. (c) Magnetic excitation spectra taken at $T$=8 K for La$_{1.82}$Sr$_{0.18}$CuO$_4$ by scanning along $(0.5,k,0)$ at several energies (solid circles). Calculated $\chi"(q,\omega)$ for the parameters $t_0=-50$ meV, $2\Delta_0=10$ meV, $J_0=55$ meV, $\Gamma_0=4$ and $\Gamma_h=20$ meV are also shown by the solid lines.

Fig. 11   Carrier concentration dependence of the quasi particle broadening at $|\varepsilon|>2\Delta_0$ ($\Gamma_h$). Open squares, and filled and open circles represent the data obtained for YBa$_2$Cu$_3$O$_y$, La$_{2-x}$Sr$_x$CuO$_4$ and La$_{1.48}$Nd$_{0.4}$Sr$_{0.12}$CuO$_4$, respectively.

Fig. 12   Magnetic excitation spectra $\chi"(q,\omega)$ calculated for Nd$_{1.85}$Ce$_{0.15}$CuO$_4$ at 5 K by using eq. (1) for the effective band parameters $t_0(=-40$ meV), $t_1=-t_0/6$, $t_2=t_0/5$ and $U=0$, are shown along [10] direction in the 2–dimensional reciprocal space for $\omega=5$ and 10 meV. Inset shows the fitted results to the experimental data reported in ref. 23.

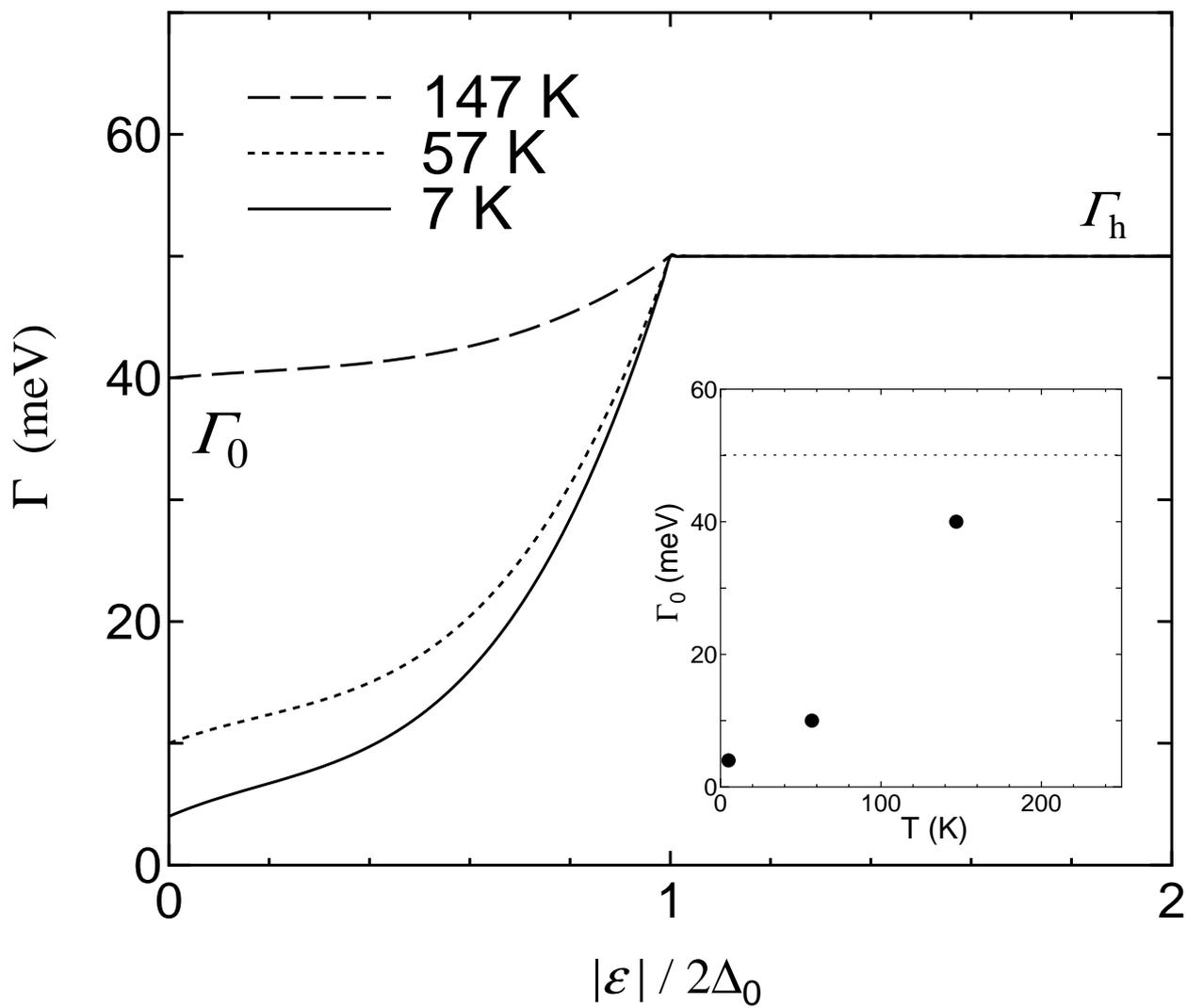

Figure 1
M. Ito et al.

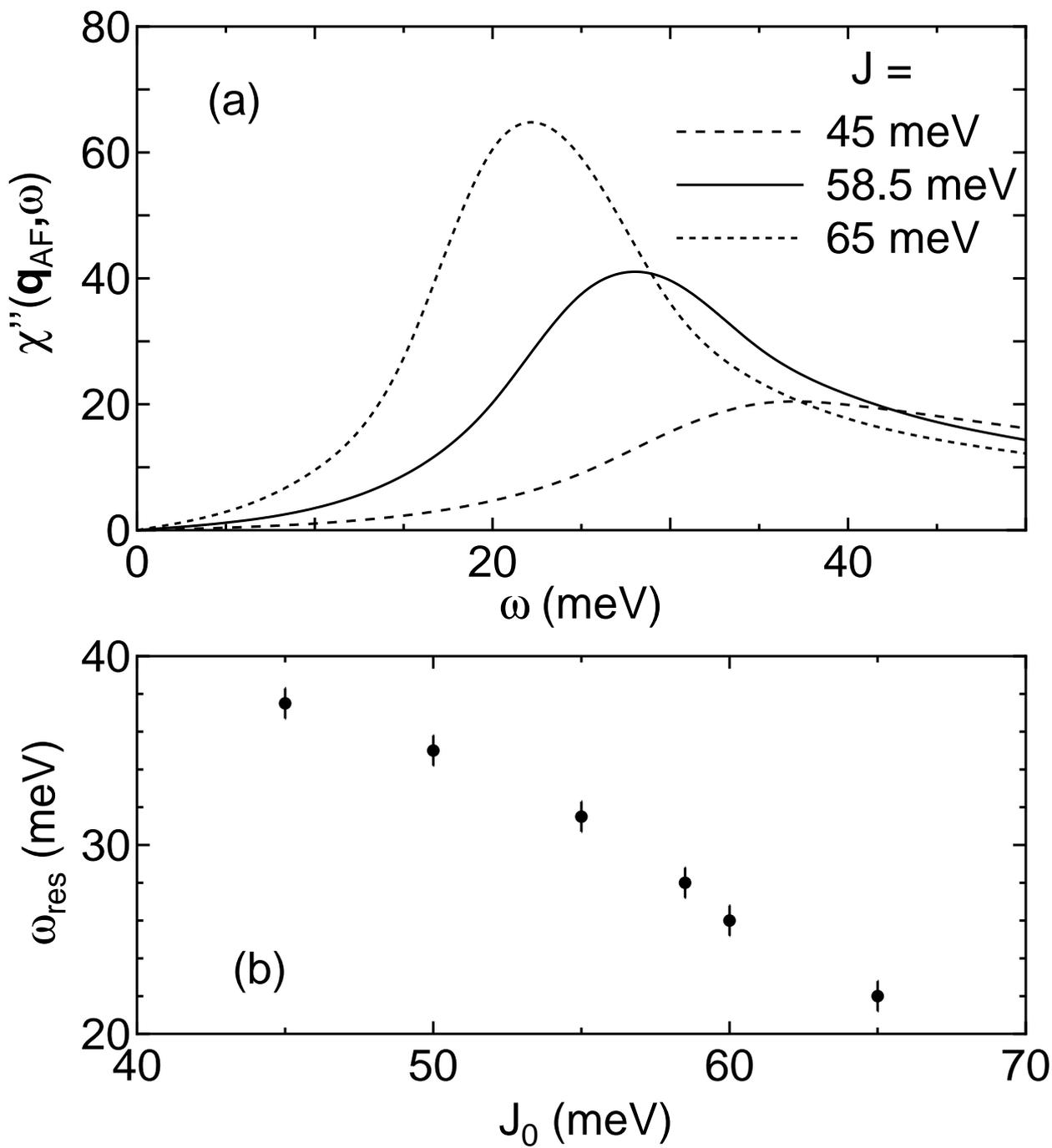

Figure 2
M. Ito et al.

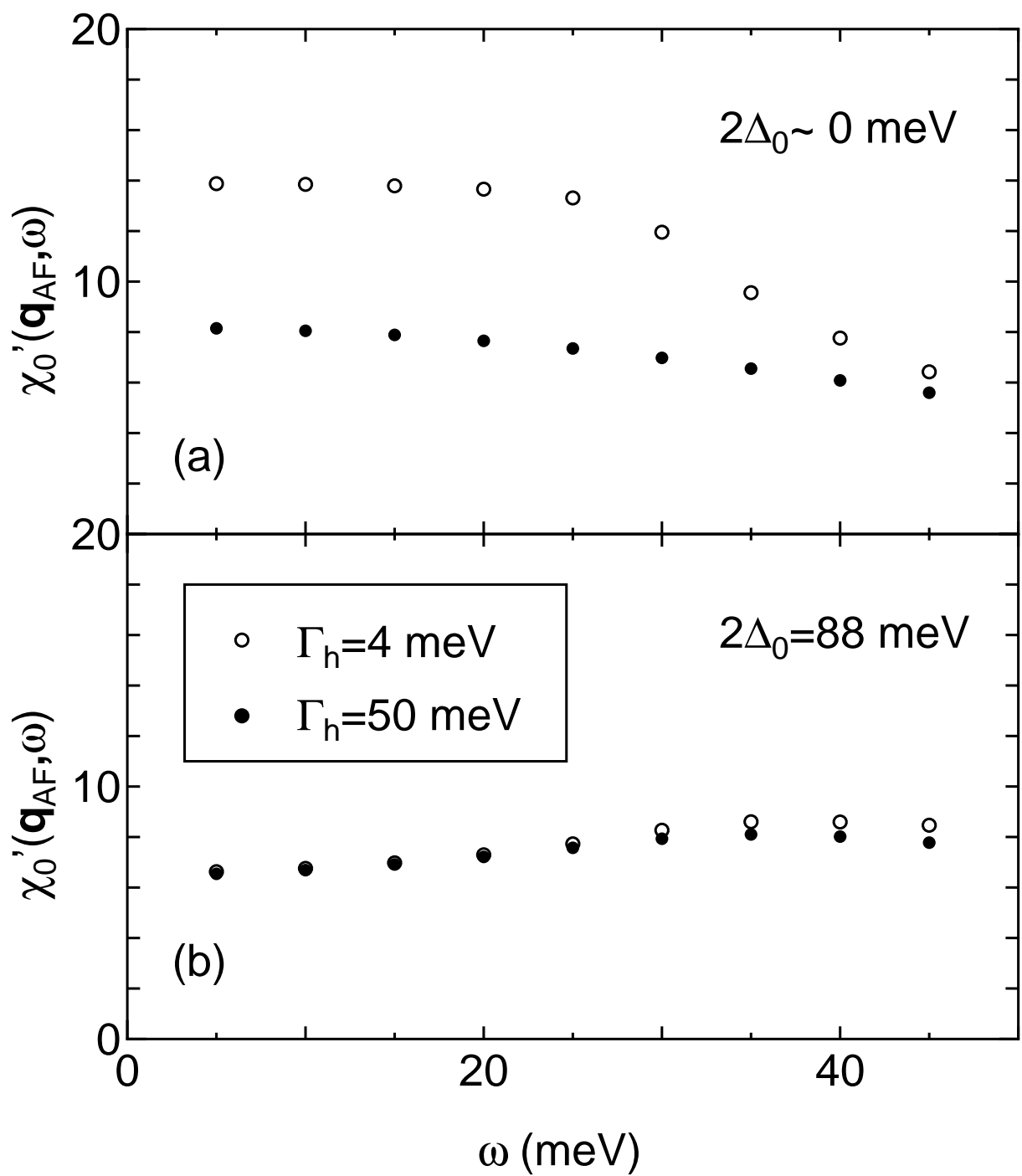

Figure 3
M. Ito et al.

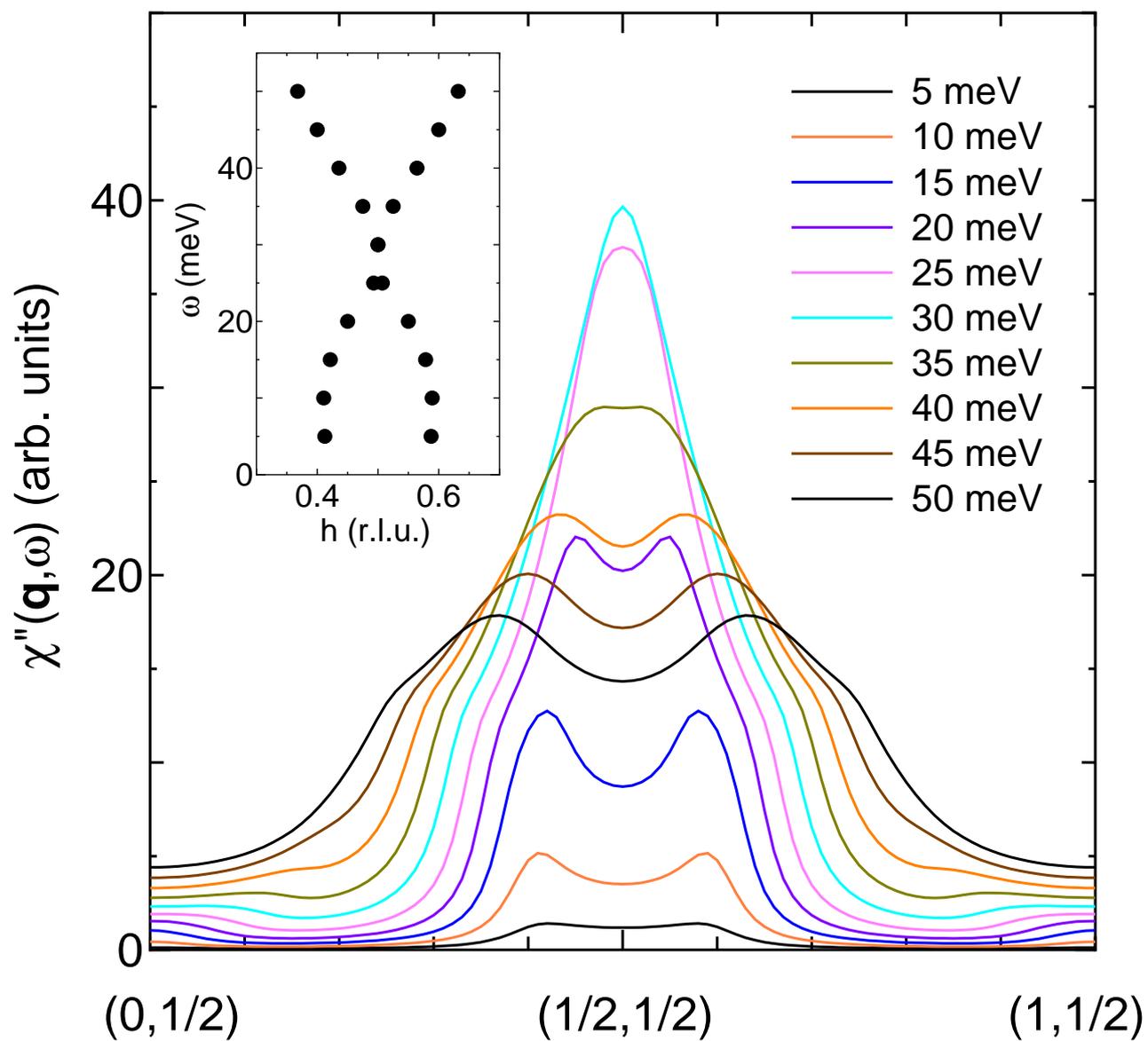

Figure 4
M. Ito et al.

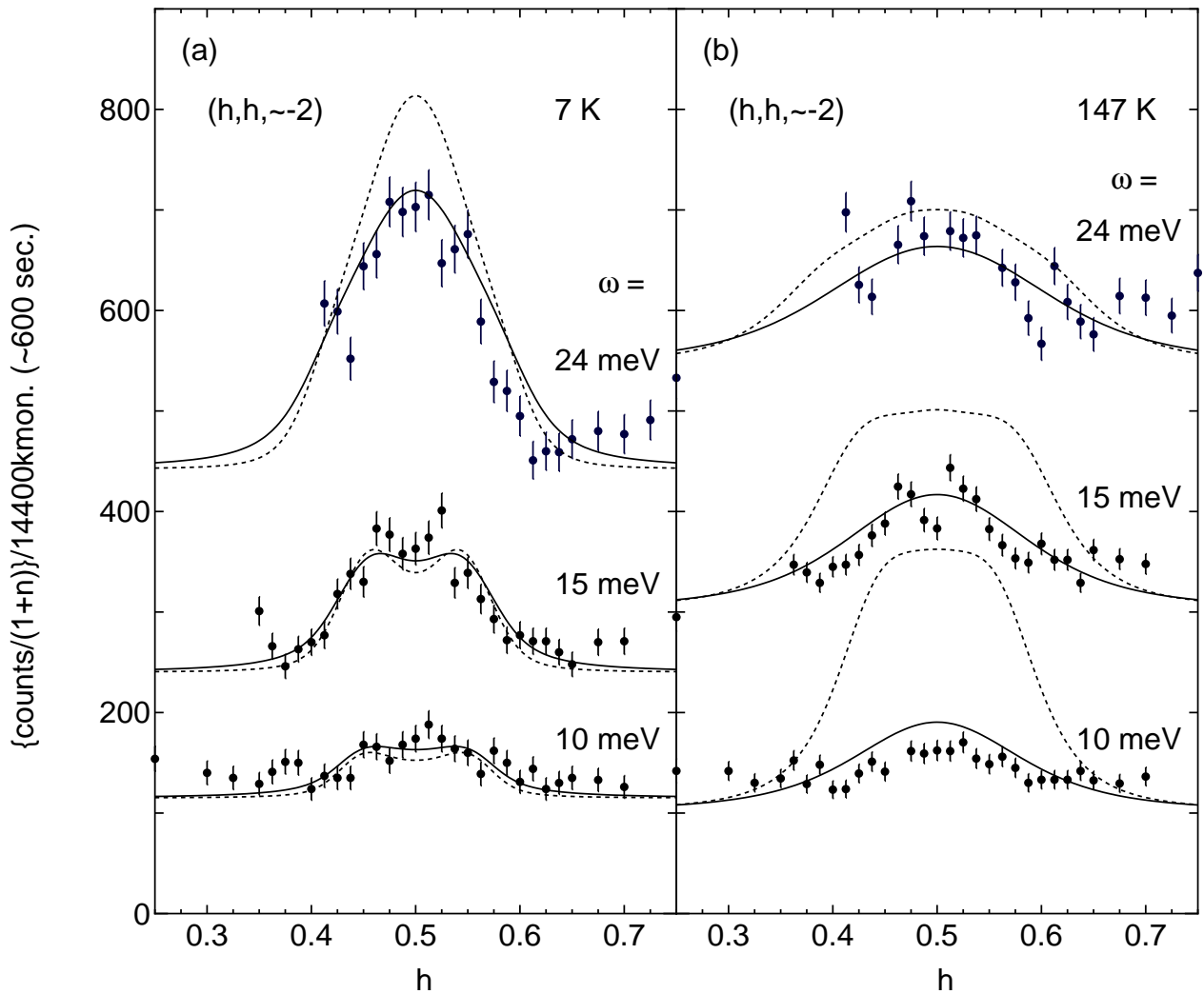

Figure 5
M. Ito et al.

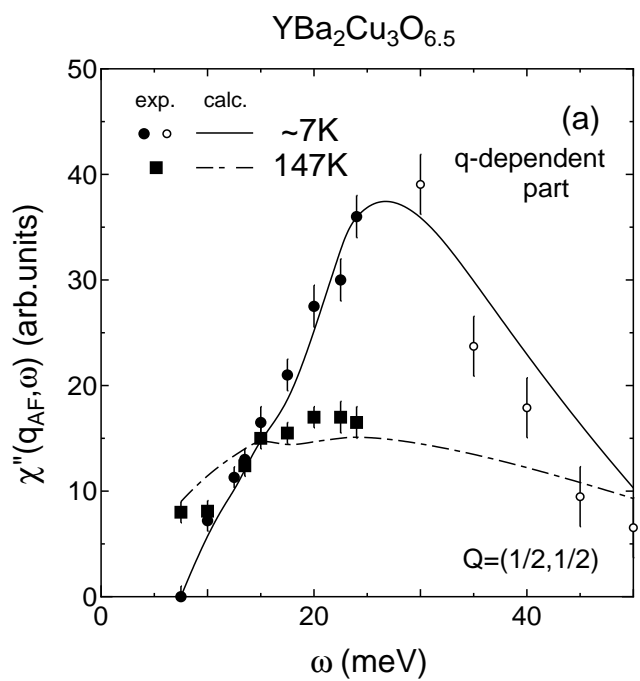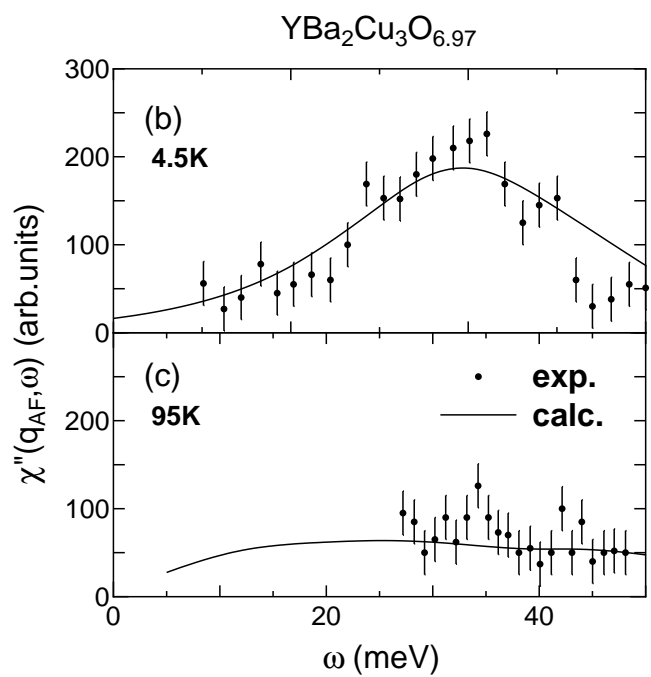

Figure 6
M. Ito et al.

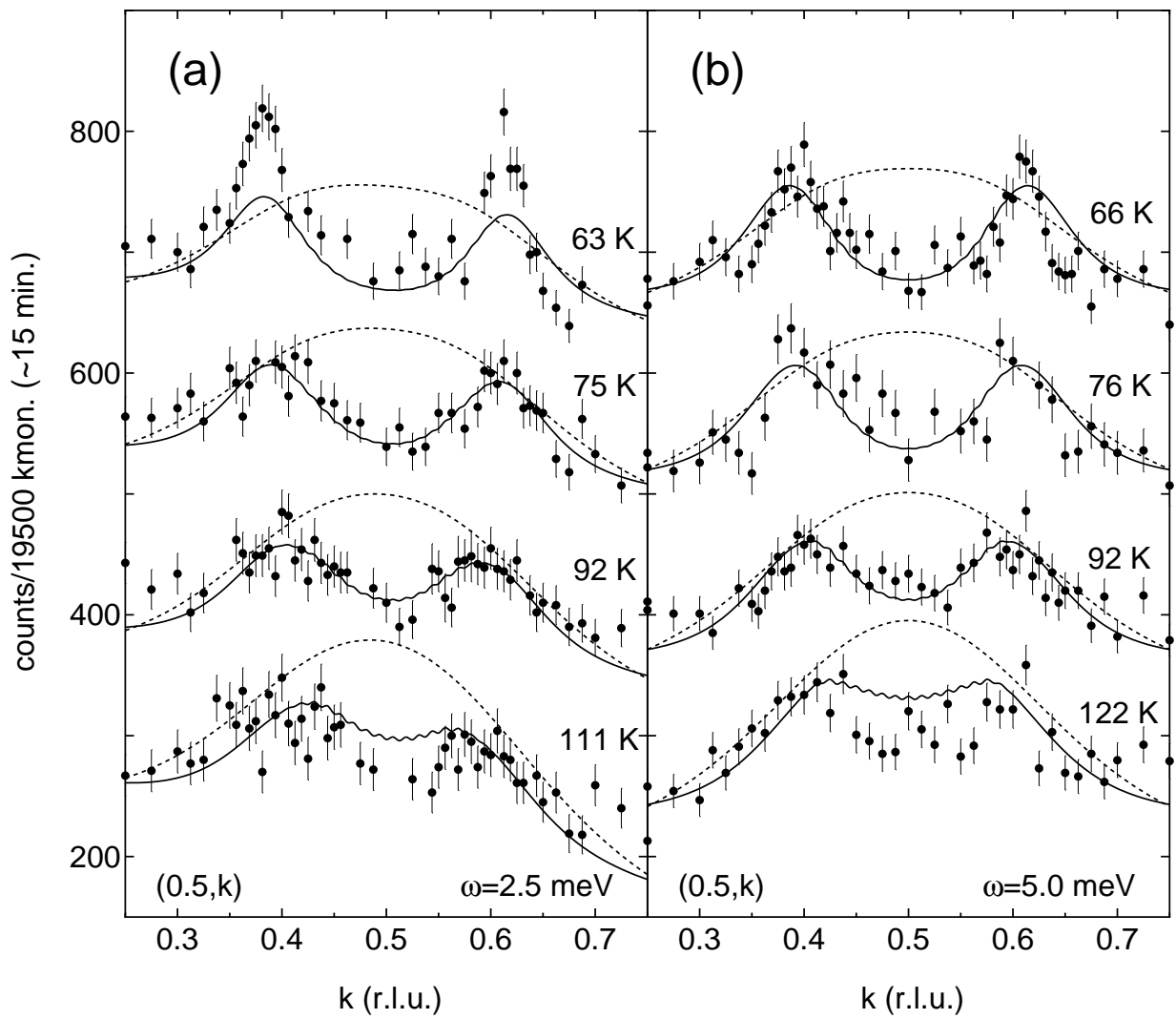

Figure 7
M. Ito et al.

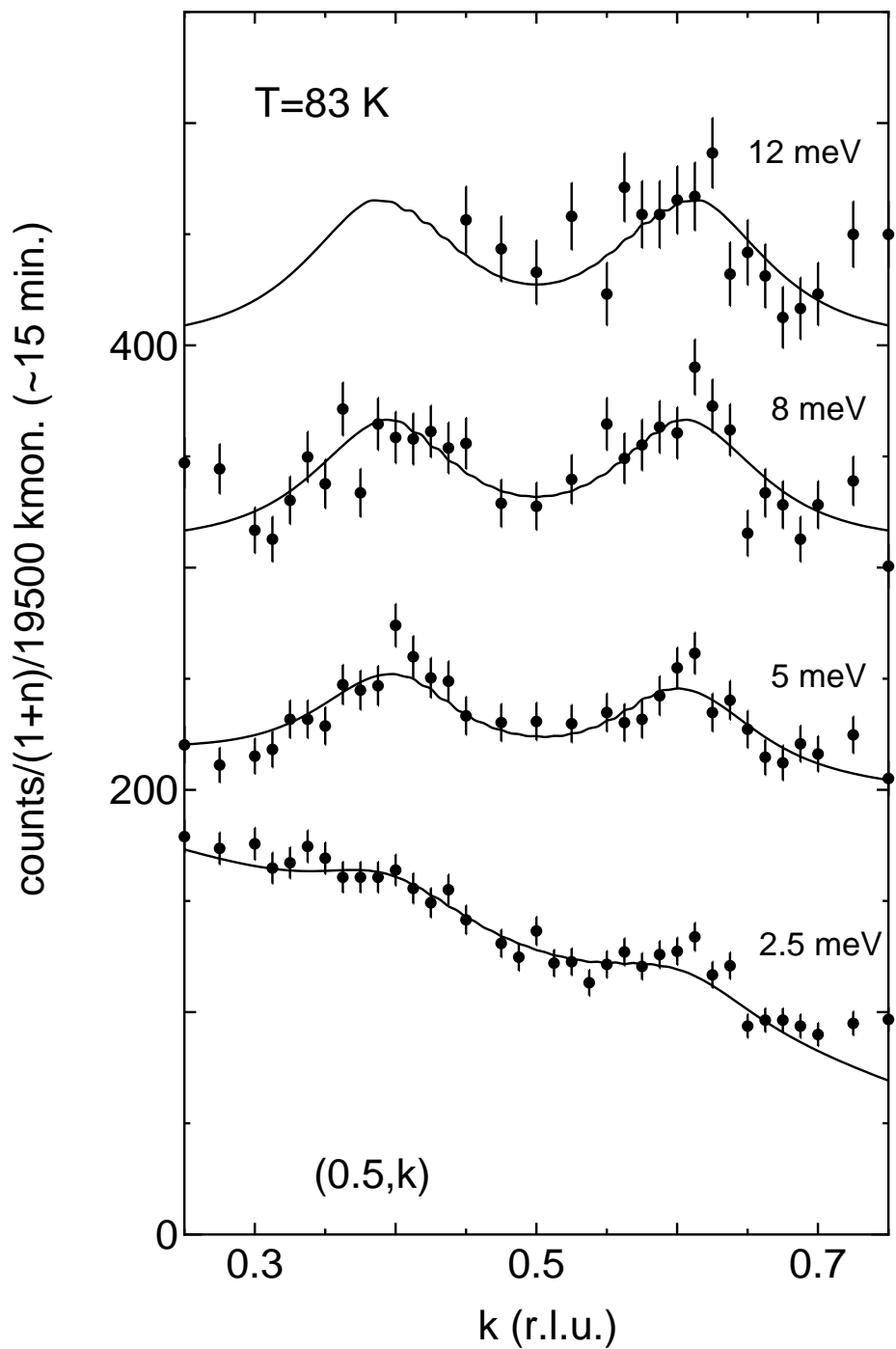

Figure 8
M. Ito et al.

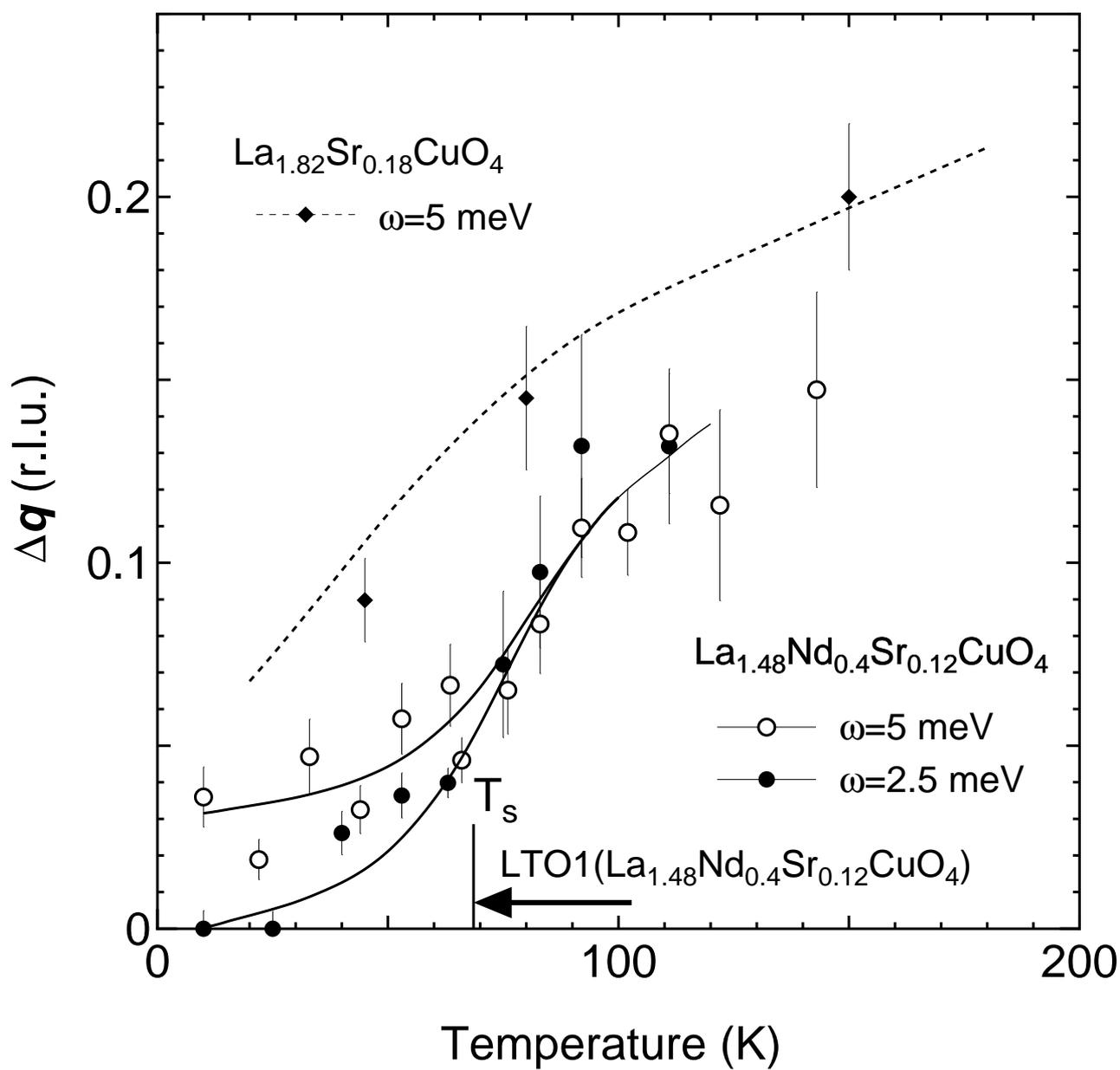

Figure 9
M. Ito et al.

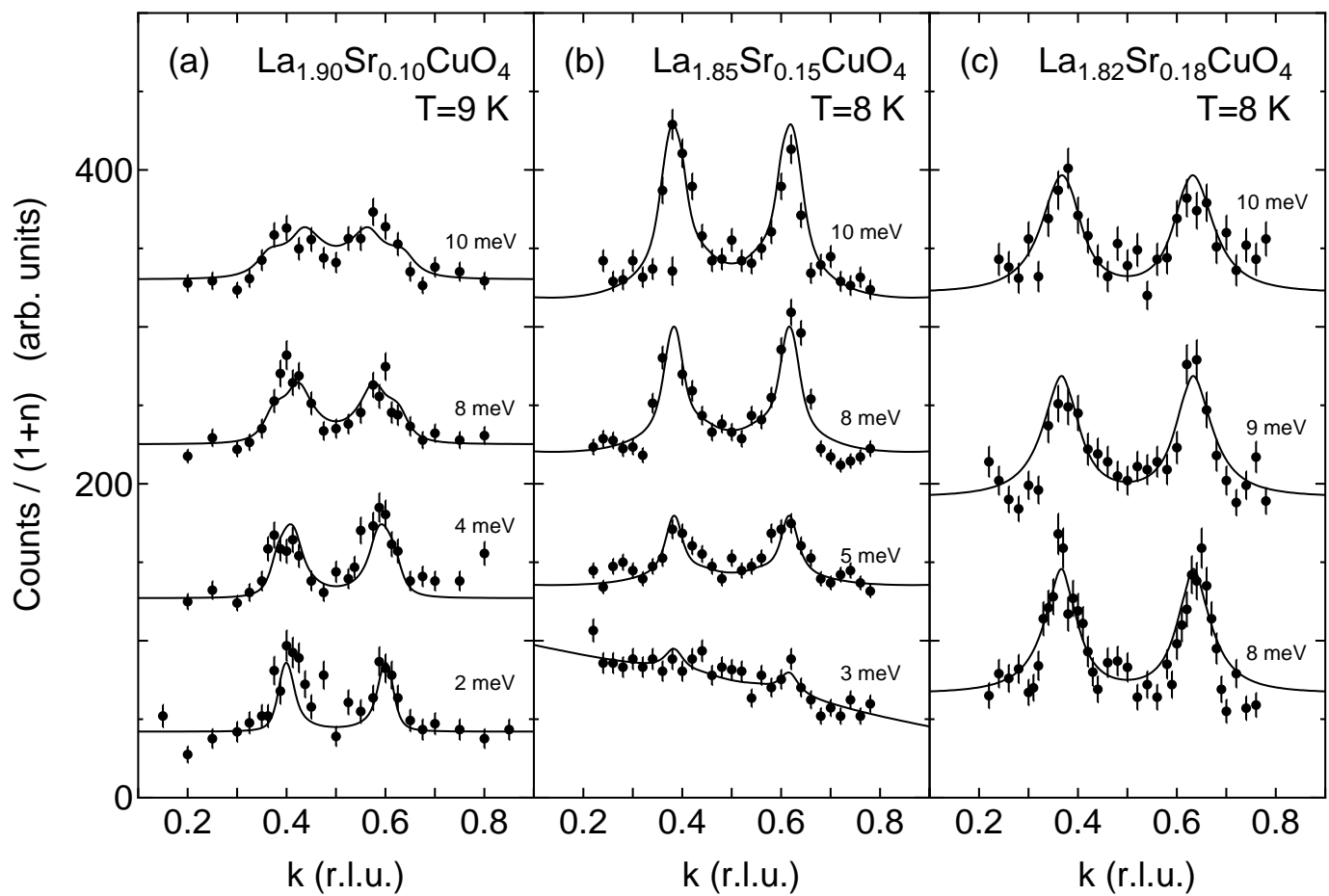

Figure 10
M. Ito et al.

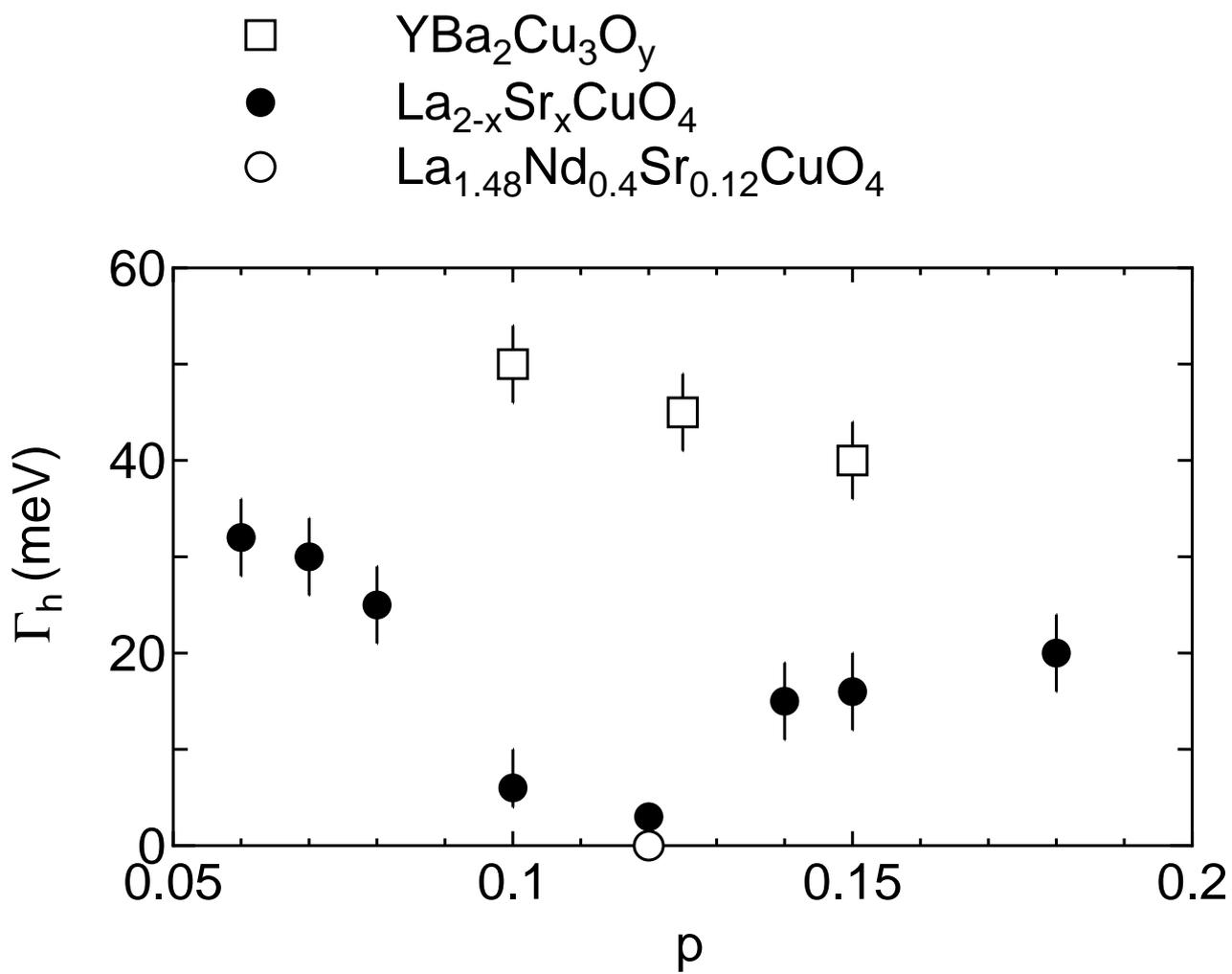

Figure 11
M. Ito et al.

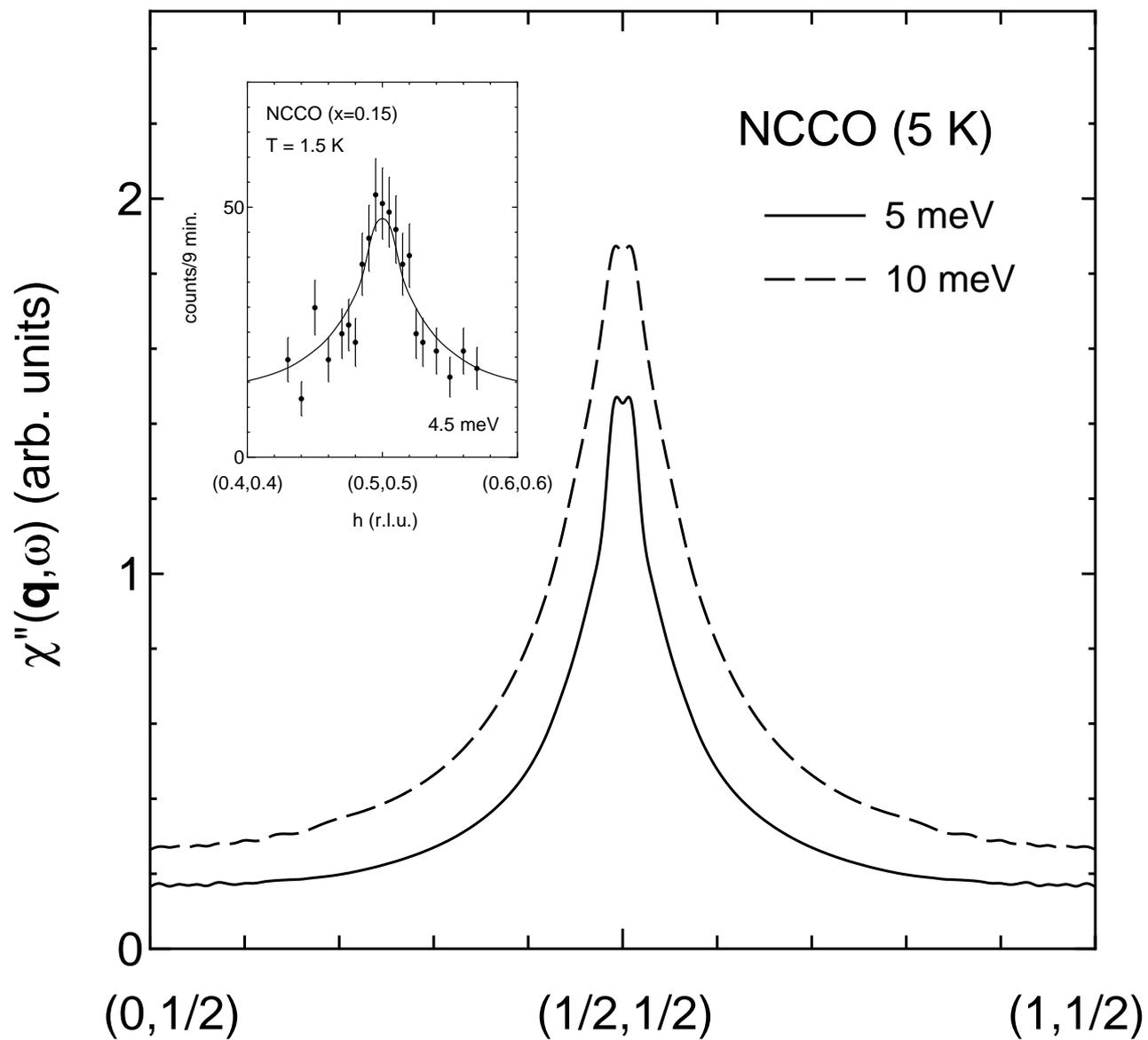

Figure 12
M. Ito et al.